\documentclass{svjour2}                    
\smartqed  
\usepackage{graphicx}
\usepackage{epsfig}
\journalname{Foundations of Physics}
\begin{document}

\title{Proposed Test of Relative Phase as Hidden Variable in Quantum Mechanics
}


\author{Steven Peil 
}


\institute{S. Peil \at
              United States Naval Observatory \\
              3450 Massachusetts Ave., NW\\
              Washington, DC  20392  USA\\
              Tel.: \\
              Fax: \\
              \email{steven.peil@usno.navy.mil}           
}

\date{Received: date / Accepted: date}

\maketitle

\begin{abstract}
We consider the possibility that the relative phase in quantum mechanics plays a role
in determining measurement outcome and could therefore serve as a ``hidden'' variable.
The Born rule for measurement equates the probability for a given outcome with the
absolute square of the coefficient of the basis state, which by design removes the
relative phase from the formulation. The value of this phase at the moment of
measurement naturally averages out in an ensemble, which would prevent any dependence
from being observed, and we show that conventional frequency-spectroscopy measurements
on discrete quantum systems cannot be imposed at a specific phase due to a
straightforward uncertainty relation. We lay out general conditions for imposing
measurements at a specific value of the relative phase so that the possibility of its
role as a hidden variable can be tested, and we discuss implementation for the
specific case of an atomic two-state system with laser-induced fluorescence for
measurement.

\end{abstract}

\section{Introduction}
\label{intro}

\subsection{Problems with Quantum Measurement}

Quantum theory prescribes probabilities for outcomes of measurements performed on
microscopic systems.  Coherent evolution of a system is governed by the Schrodinger
equation, allowing determination of the time-dependent amplitudes and phases of the
states characterizing the system; the amplitudes of the measurement-basis states at
the time of measurement determine the probabilities for the corresponding outcomes.
While this formulation has been remarkably successful, there are issues associated
with quantum measurement that have existed from the development of the
theory~\cite{wheeler}\cite{adler}.

First, the fact that the theory is probabilistic and is unable to determine the
outcome of a measurement on an individual system has been troubling to some, leading
to suggestions that the quantum description of a system is incomplete and that there
could be ``hidden variables'' that might determine measurement outcome. Second, there
is no known underlying origin for the Born rule determining measurement probabilities;
it is a postulate of the theory. This is unlike statistical mechanics, in which
probabilities can be explained in terms of deterministic classical behavior. Finally,
the unitary evolution governed by the Schrodinger equation cannot result in the
observed outcome of a measurement; at what level of microscopic description the
non-unitary measurement process enters is ill-defined.  This is commonly known as the
(quantum) measurement problem.

Despite claims that the success of quantum mechanics implies that these foundational
problems are of little consequence, many have emphasized the importance of resolving
these issues, either by re-interpretation or modification of the theory. Modifications
required to address these issues could have significant implications on efforts to
unify gravity and quantum mechanics~\cite{adler}\cite{smolin}.

\subsection{Testable Modifications to Quantum Mechanics}

Some interpretations of quantum mechanics, such as the many-worlds and
consistent-histories formulations, try to address the quantum measurement problem and
the origin of the Born rule~\cite{MW}\cite{histories}. But different interpretations
of quantum mechanics make identical predictions, so these alternate theories must be
considered untestable.  On the other hand, models attempting to resolve the
measurement problem have been proposed that introduce new features and make
predictions that differ from conventional quantum mechanics, making the proposed
theories subject to verification. An example is the model of continuous spontaneous
localization, which attempts to treat wave function collapse with a stochastic noise
term added to the Schrodinger equation and which makes predictions that should be put
to the test by experiments attempting to demonstrate coherence on systems of
increasing size and complexity~\cite{CSL1}\cite{CSL2}.  In these efforts to address
the measurement problem, the probabilistic nature of quantum mechanics is considered a
postulate.

Similarly, for attempts to address nondeterminism in quantum mechanics,
interpretations or even putative modifications which make no predictions different
from conventional theory are untestable.  The deBroglie-Bohm pilot-wave theory is
subject to this criticism~\cite{dBB}.  Bell's theorem and related work enable tests of
general classes of hidden-variable theories.  These tests contrast predictions of
quantum mechanics with those of theories including local hidden variables, at least as
applied to measurements on individual members of correlated systems.  All experiments
so far indicate that any hidden-variables must be nonlocal.  While tests of general
properties of hidden-variables theories have been fruitful, we are unaware of any
experimental tests for a specific, proposed hidden variable.

\subsection{Relative Phase and Measurement Outcome}

The only prediction of measurement outcomes possible in quantum mechanics is the
likelihood for a certain outcome given by the square of the coefficient of the
measurement-basis state.  The relationship between measurement outcome and probability
amplitudes dictated by the Born rule has been validated implicitly due to its ubiquity
in quantum theory, but searches for dependence of measurement outcome on other
parameters have been lacking.

For a two-state system, normalization constrains the measurement probabilities and
only one independent parameter is involved in predicting measurement outcome, the
population difference between the basis states.  The relative phase is the only other
independent parameter characterizing a two-state system, and Born's rule for deriving
measurement probabilities eliminates the phase from the formulation by construction.
Here we consider the possibility of the relative phase of a two-state system playing a
role in determining measurement outcome and therefore acting as a hidden variable.

\section{Relative Phase as Hidden Variable}

The possibility of the relative phase as a hidden variable has not been ruled out
explicitly--no direct tests have been carried out, nor implicitly--it is naturally
averaged over in the measurement process, eliminating the signature of any possible
role in measurement outcome.  Yet, specifically engineering the measurement process so
that it occurs at a specific value of the relative phase enables a search for a role
as a hidden variable~\cite{mod}.

\subsection{Bell's Theorem}

The traditional framework for considering hidden variables is based on the work of
John Bell, who showed that quantum mechanics predicts measurements made on entangled
particles can exhibit correlations that can not be explained with local hidden
variables~\cite{bell}. Bell-type tests are typically applied to two particles
entangled in a state such as
\begin{equation}
|\Psi\rangle=\frac{1}{\sqrt{2}}\left(|0\rangle|1\rangle+e^{i\phi}|1\rangle|0\rangle\right).
\label{e.singlet}
\end{equation}
This could represent, for example, two spin-1/2 systems in a singlet state for
$\phi=\pi$. Measurements on a particle are made in one of several bases, and the
measurements applied to each system are space-like separated, so that the basis for
the measurement made on system 1 can not be communicated to the measurer of system 2.
The quantum correlations associated with the measurement outcomes therefore can not be
attributed to local hidden variables.  All Bell-type tests to date are consistent with
the results of traditional quantum mechanics and imply that any hidden variables that
could exist must be nonlocal.

The relative phase of a two-state system is not ruled out as a possible hidden
variable by Bell's theorem because it is fundamentally not a local variable; if the
system is in a superposition of widely separated states, the relative phase has to be
considered nonlocal. Additionally, the framework for Bell's theorem does not directly
apply.  The relative phase is a property of a superposition state, and entanglement is
not necessarily involved.  For a state like that in Eq.~\ref{e.singlet}, testing the
possibility of phase as a hidden variable requires looking for a dependence of
measurement outcome versus the value of $\phi$ at the moment of measurement.  However,
because the measurement process significantly affects the relative phase, the value at
the moment of measurement may be very different than the value in the original state.

\subsection{Existing Experiments}

Searching for a dependence of measurement outcome on the value of the relative phase
at which measurement occurs is difficult. The value of the phase is affected by the
measurement process, to the extent that the relative phase at the moment of
measurement naturally averages out over an ensemble. A special technique for forcing
measurement to occur at a specific, reproducible phase is required. These points are
discussed in detail in Section~\ref{s.phase-specific}, where we introduce a general,
quantitative model of measurement on a two-state system; we emphasize the abruptness
of decoherence compared to the phase evolution of the system, demonstrating that
measurement at a specific value of phase is meaningful; we evaluate the phase
evolution of the system introduced during the measurement process, showing via an
uncertainty relation that the value of the relative phase at the moment of measurement
naturally averages out over an ensemble; and we present measurement strategies that
circumvent this uncertainty relation and impose measurement at a specific phase.

\subsection{Relative Phase in Standard Quantum Mechanics}

According to standard quantum mechanics, the relative phase of a superposition state
has no effect on measurement outcome; the Born rule for deducing measurement
probabilities eliminates the relative phase by construction through the squaring of
the complex probability amplitudes.  Rather, the phase serves as a record of coherent
evolution, and it plays an important role in atomic clocks and other atomic
interferometers.  In clocks that rely on spectroscopy in which the atomic transition
is probed coherently, the clock is designed as an
interferometer~\cite{interferometer}.  The first interaction, or beam splitter,
creates a population difference in the atomic sample and initiates a period of
coherent phase evolution. The value of the relative phase after this evolution gets
``locked in'' by generating a phase-dependent population difference at the second beam
splitter. This population difference gives the information on the atomic frequency
required for clock operation.

The actual measurement of the population in the two clock states after the second beam
splitter destroys the coherence in the system.  For any method of state readout
employed in atomic clocks, the measurement process naturally perturbs the relative
phase to the extent that the measurement can not be considered to occur at a specific
value of the phase. (In fact, for atomic clocks, the phase precession is typically
fast enough that the duration of the measurement process is long compared to a cycle
of phase evolution.) It is for this type of measurement that we are interested in
considering if there is a dependence on the relative phase.

\subsection{Examples of Phase-Dependent Measurement Outcome}

If measurement outcome for a two-state system were to depend on the relative phase,
the implication would be that the Born rule would be an approximation that is only
correct when measurements are averaged over the relative phase.

Classical and semi-classical examples where measurement outcome is phase-dependent,
though in a way that is subtle or difficult to implement, are perhaps suggestive.  The
strength of the electric field of a classical optical signal varies with a phase that
evolves at a rate that can be on order of $10^{15}$~Hz. Directly measuring the
phase-sensitive field strength is impractical with optoelectronic devices, yet the
phase manifests itself much more straightforwardly through interference effects. In
the semiclassical vector model for adding two quantum angular momenta, component
momenta $\mathbf{j_i}$ are portrayed as precessing about the resultant
$\mathbf{J}$~\cite{herzberg}. For cases where there are final states with $m_J=0$, the
projections of the $\mathbf{j_i}$ along the quantization axis are always opposite in
sign and precess at a rate proportional to the interaction strength (faster for the
case representing the triplet state,
$|\uparrow\downarrow\rangle+|\downarrow\uparrow\rangle$, associated with larger
$\mathbf{J}$, than for the singlet state
$|\uparrow\downarrow\rangle-|\downarrow\uparrow\rangle$, associated with smaller
$\mathbf{J}$). Measuring the projections of the $\mathbf{j_i}$ implies interrupting
this precession, leaving the system in a configuration representing
$|\uparrow\downarrow \rangle$ or $|\downarrow\uparrow \rangle$, depending on the phase
when measured.

\section{Phase-Specific Quantum Measurement}
\label{s.phase-specific}

\subsection{General Measurement Model}

A general, two-state quantum system can be characterized with just two independent
parameters, the population difference and the relative phase between the two states.
The state vector $|\psi\rangle$ can be written as
\begin{equation}
|\psi\rangle = \alpha |0\rangle + e^{i\phi} \sqrt{1-\alpha^2} |1\rangle,
\label{e.superposition}
\end{equation}
where the amplitude $\alpha$ and relative phase $\phi$ are real, and $|0\rangle,
|1\rangle$ are orthogonal basis states.  For an isolated system, with energy
difference $E$ between the higher energy state $|1\rangle$ and ground state
$|0\rangle$, a superposition state evolves with a relative phase $\phi=
\frac{E}{\hbar} t$ and a period $\tau_\phi=\frac{2\pi \hbar}{E}$. The measurement
postulate of quantum mechanics states that measuring the system in the
$\{|0\rangle,|1\rangle\}$basis will result in state $|0\rangle$ with a probability of
$\alpha^2$ (and state $|1\rangle$ with probability $1-\alpha^2$). These measurement
probabilities derive from the Born rule, $P_x=| \langle x | \psi \rangle |^2$, an
expression which dismisses the phase factor $e^{i\phi}$ between the basis states that
represent potential measurement outcomes.

\subsection{System, Reservoir Timescales}

In order to carry out a measurement, the system must be coupled to a macroscopic
reservoir that introduces irreversible evolution by damping the energy of state
$|1\rangle$ at a rate $\Gamma_1$.  A quantum fluctuation can introduce correlations
between the system and reservoir, which decay in a time $\tau_c$~\cite{cohen-tann}. In
terms of measurement on the two-state system, the vanishing correlations mark the
point of irreversibility. The correlation time is roughly the inverse bandwidth of the
reservoir and is very short for a macroscopic reservoir. In most cases, $\tau_c$ is
much shorter than any time for the quantum system to evolve, so that the measurement
process takes place over a very short interval of time and a very small range of
relative phases $\Delta \phi$, which we can take to be zero. This amounts to the
Markov approximation for the reservoir.

A quantum fluctuation corresponds to measurement on the coherent two-state system via
the presence or absence of spontaneous emission from the $|1\rangle$ to $|0\rangle$
transition~\cite{null}. The average time needed to acquire information from the system
for a measurement is the lifetime of state $|1\rangle$, $1/\Gamma_1$, which we call
the measurement time $\tau_m$~\cite{meas_time}. Once the system and reservoir are
coupled (some amplitude in $|1\rangle$), the probability of a quantum fluctuation
resulting in measurement at time $t$ is $P(t)\propto e^{-t/\tau_m}$; beyond this, the
specific time for a fluctuation to occur is unpredictable (Fig.~\ref{f.reservoir}(a)).
A measurement can be applied at a specific $\phi$, then, in a system in which $\tau_m$
is short compared to the period $\tau_\phi$.

\subsection{Measurement at Specific Value of $\phi$}

For systems where the phase evolution is very fast, such as in an atomic clock, this
requirement on the measurement time is stringent. In fact, this condition is {\em
never} satisfied, no matter how slow the phase evolution of the system, when the
method used to differentiate the two states relies on resolving their frequency
difference. The Fourier-transform limit for resolving a frequency difference $\Delta
\nu$ requires measuring for an interval of time $\Delta t \geq 1/\Delta \nu$. 
Since the energy difference between the states is $E=2\pi \hbar \Delta \nu$, during
this time the relative phase of a superposition state evolves by an amount
\begin{equation}
\frac{1}{\hbar} \int_0^{\Delta t}  E~dt' = 2 \pi \Delta \nu \Delta t \geq 2 \pi. 
\end{equation}
So the measurement time required to resolve two states spectroscopically is always on
order of $\tau_\phi$, and measurement can occur at any value of $\phi$ between 0 and
2$\pi$ radians. (The only phase specificity in this case is due to the exponential
probability distribution for a measurement to occur.)  We know of no measurements on
two-state systems, including atomic clocks or specific qubits in quantum information,
in which the phase of the state at the moment of measurement is not averaged out over
an ensemble.

\begin{figure}
  \includegraphics[width=0.75\textwidth]{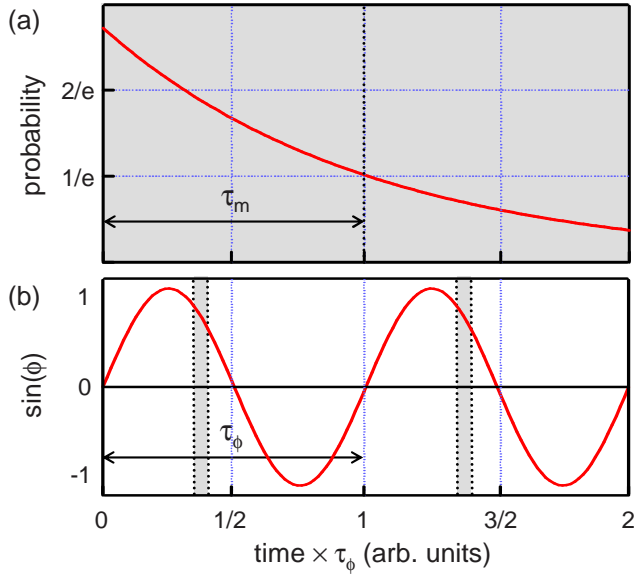}
\caption{(color online.) (a)~Probability of quantum fluctuation that entangles system
and reservoir versus time after interaction is turned on. We call the average time for
this distribution the measurement time $\tau_m$. Measurement can occur at any time
within the shaded area. (b)~Sine of relative phase $\phi$ versus time. Measurement can
be applied at a specific $\phi=\frac{E}{\hbar} t$ by applying brief measurement pulses
in synch with the phase evolution.  These pulses--represented by the shaded
intervals--are the only time the system-reservoir interaction is on and measurement is
possible. The specific plots (a) and (b) correspond to the Fourier transform limit for
a spectroscopic measurement, for which $\tau_m=\tau_\phi$.}
\label{f.reservoir}       
\end{figure}

To get around this, a spectroscopic measurement can be applied in which the coupling
to the reservoir is on only during a specific, narrow range of values of $\phi$, as
illustrated in Fig.~\ref{f.reservoir}(b). This brief interaction, much shorter than
$\tau_m$, is unlikely to result in measurement, but it can be repeated periodically,
in synch with the coherent evolution of the phase, and after many such interactions
the probability of measurement can approach unity. A second approach to realizing
phase-selective measurement is to rely on a different method of state discrimination,
such as using selection rules, in which case there is no Fourier-transform limit on
the measurement time, allowing it to be much shorter than $\tau_\phi$.

\section{Application to Atomic System}

We will discuss these examples in some detail for an atomic system. The two-state
system can be realized with two long-lived electronic ground states. Different
hyperfine levels typically have a frequency difference that is too large to satisfy
$\tau_m \ll \tau_\phi$. However, Zeeman levels within a hyperfine manifold are
degenerate at zero magnetic field and have a splitting that can be tuned with field.
For low fields, the splitting between adjacent Zeeman sublevels is $\Delta \nu =g
\mu_B B$, where $g$ is the Lande $g-$factor for the states and $\mu_B$ the Bohr
magneton.  For rubidium ($^{87}$Rb), for example, $\Delta \nu = 0.7~$MHz/G for the
total angular momentum $F=2$ ground-state manifold.

\begin{figure}
\includegraphics[width=0.75\textwidth]{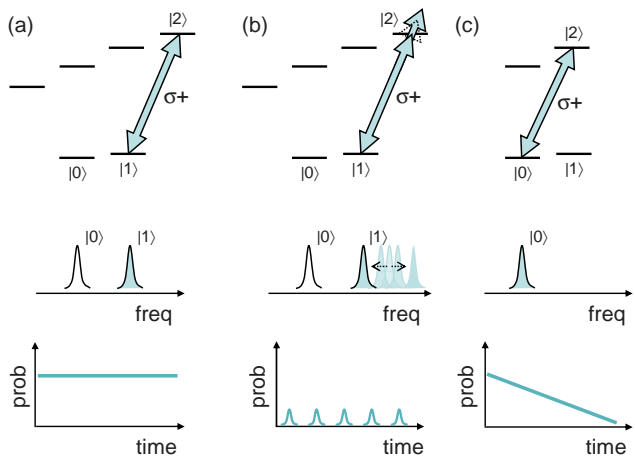}
\caption{(color online.) Energy level diagrams showing $\sigma +$ transitions between
two-state system and state $|2\rangle$ (top), illustration of frequency of $\sigma +$
transitions involving $|0\rangle$ and $|1\rangle$ (middle), and illustration of
probability of scattering a photon versus time (bottom) for three scenarios.
(a)~Standard atomic-state detection technique relies on tuning a detection laser to be
resonant with a specific transition. Circularly polarized detection light drives a
closed $\sigma +$ transition resulting in a constant average probability for an atom
in state $|1\rangle$ to scatter a photon. (b)~In order to impose a measurement at a
specific $\phi$, the frequency of the detection laser can be modulated synchronously
with the phase evolution of the superposition state.  By also adjusting the phase of
the drive as discussed in the text, the probability of scattering a photon becomes a
series of pulses. (c)~For a detection transition between levels with the same angular
momentum, selection rules can be used for state discrimination.  Here, there are no
cycling transitions, and the probability to scatter a photon decreases in time.}
\label{f.atomlevels}
\end{figure}

In general, measurement is not carried out by looking for the decay $|1\rangle
\rightarrow |0\rangle$ as discussed above. For a naturally well isolated system,
introduction of a third state, $|2\rangle$, which is more strongly coupled to its
environment, enables more efficient and sensitive detection. For an atomic system,
this third state is typically separated from the long-lived states by an optical
frequency. Measurement in the $\left\{|0\rangle,|1\rangle\right\}$ basis can be
carried out by applying laser light that is tuned to the $|1\rangle$ to $|2\rangle$
transition. The optical drive coherently transfers amplitude to $|2\rangle$, which
decays via spontaneous emission that can be detected as an indicator of population in
state $|1\rangle$; the absence of spontaneous emission indicates population in state
$|0\rangle$.  The emission process can be repeated many times if the transition is
closed, {\em i.e\@.} if state $|2\rangle$ can only decay to state $|1\rangle$; this is
illustrated in Fig.~\ref{f.atomlevels}(a) for the case of circularly polarized light
driving a $\sigma +$ transition. Here, the measurement time for the $\left\{
|0\rangle,|1\rangle \right\}$ basis is the average time to scatter a photon on the
$|1\rangle \leftrightarrow |2\rangle$ transition, $\tau_m=1/\left( P_2 \Gamma_2
\right)$, where $\Gamma_2$ is the spontaneous emission rate and $P_2$ the absolute
square of the amplitude of state $|2\rangle$. The correlation time for the vacuum is
less than a period of the optical frequency, $\tau_c < 1/\nu_{21}$, so the condition
$\tau_c \ll \tau_m$ is easily satisfied~\cite{cohen-tann}.

\subsection{Measurement Pulses}

The first approach to engineering a phase-specific measurement requires generation of
measurement ``pulses,'' brief intervals of time during which a quantum fluctuation
resulting in measurement can occur.  For our example, these pulses will be times when
the detection laser can cause the atom to scatter a photon. Short intervals of
resonant light can be created by frequency modulating the laser so that the detuning
from the atomic transition is close to zero only for a small fraction of the
modulation cycle (see illustration in Fig.~\ref{f.atomlevels}(b)). We consider the
detection laser frequency $\nu_d$ centered above the atomic resonance, $\nu_{21}$,
away from other nearby $\sigma +$ transitions, and sinusoidally modulated so that it
is resonant with the detection transition at an extremum.  In terms of the detuning
$\delta=2\pi(\nu_d-\nu_{21})$,
\begin{equation}
\delta=2\pi \nu_{\rm off}\left(1+\cos(2\pi\nu_{\rm mod} t) \right),
\end{equation}
where the modulation frequency, $\nu_{\rm mod}$, must match the frequency of the
two-state system, $\nu_{10}$, to keep the detection pulses in synch with the phase
evolution, and the offset frequency $\nu_{\rm off}$ then determines the fraction of
the modulation cycle during which the laser can cause a photon to be scattered.

\begin{figure}
\includegraphics[width=0.75\textwidth]{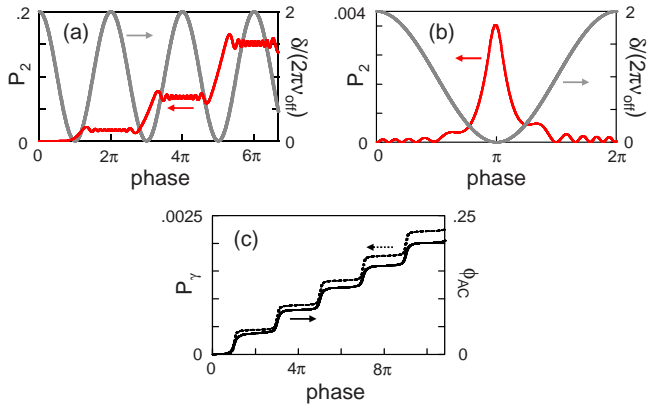}
\caption{(color online.) Integration of the optical Bloch equations yields the
population in the excited state $|2\rangle$, $P_2$, due to the modulated detection
laser. The calculations here are for the parameters $\{2\pi\nu_{\rm off},2\pi\nu_{\rm
mod},\Omega_R,\Gamma_2\}=\{100,10,2,1\}$. (a) Plots of $P_2$ (red curve, left axis)
and $\delta$ (grey curve, right axis) versus phase of modulation. $P_2$ increases with
each modulation cycle, and spontaneous emission can occur at any value of the phase of
the two-state system. ($P_2$ is calculated for $\Gamma_2=0$.) (b) Adjusting the phase
of the detection drive at resonance can reverse the population transfer in the first
half of the modulation cycle, leaving the atom in $|1\rangle$, and creating an
interval outside of which the likelihood of measurement is small. Shown in the plot is
$\delta$ (grey curve) for one modulation cycle and $P_2$ (red curve) for one
measurement pulse. (c) Plots of probability of measurement (dashed curve, left axis)
and accumulated phase on the $|1\rangle \leftrightarrow |0\rangle$ transition from the
AC Stark shift due to the detection laser (solid curve, right axis) versus phase of
modulation. For this example, on order of 2000 pulses are required to ensure that a
measurement occurs.} \label{f.pulse}
\end{figure}

The brief intervals of resonant light do not constitute measurement pulses. A
modulation cycle transfers some amplitude from $|1\rangle$ to $|2\rangle$, as shown in
Fig.~\ref{f.pulse}(a).  Because the probability of spontaneous emission is
proportional to the square of the amplitude in the excited state, spontaneous emission
can occur at any time and any phase--the only way to turn off spontaneous emission is
to make the amplitude for state $|2\rangle$ zero.

The interval of resonant light experienced by the atom during a modulation cycle can
be converted into a measurement pulse by adjusting the phase of the optical drive
halfway through the resonant interval to reverse the initial population transfer and
leave the atom in state $|1\rangle$, where no spontaneous emission can occur.  For a
constant detuning, reversing the population transfer would require a phase change of
the optical drive of $\pi$; the required change in the case of a particular frequency
modulation can be determined empirically by integration of the optical Bloch
equations. For the example in Fig.~\ref{f.pulse}(b), the phase shift required is 0.75
radians. This adjustment to the phase can be implemented with an electro-optic
modulator, which can be expected to require a time on order of 1~ns for a general
phase change~\cite{eom}.  This imposes constraints on the duration of the detection
pulse, and to make the pulse phase selective, on the period of evolution of the
two-state system $\tau_\phi$; the frequency splitting $\nu_{10}$ has to be kept at or
below about 50~MHz, which, for the rubidium example introduced earlier, corresponds to
a magnetic field of 70~G.

Figure~\ref{f.pulse}(c) shows the total probability that a measurement has occurred,
$P_\gamma$, as a function of phase of the frequency modulation. The measurement pulse
in Fig.~\ref{f.pulse}(b) corresponds to a Rabi frequency $\Omega_R$ on the $|1\rangle$
to $|2\rangle$ transition of $2\Gamma_2$, which leads to a 0.05\% probability of a
photon being scattered (measurement being made) per measurement pulse.  Because the
frequency modulation is not symmetric about the detection transition, there is a
finite AC Stark shift on the $|0\rangle$ to $|1\rangle$ transition of .05 radians per
modulation cycle.  This needs to be accounted for by modifying the modulation
frequency so that the modulation cycle and atomic coherence are synchronized.  For
this example, the modulation frequency needs to be decreased by less than 1\%.

\subsection{Selection Rules}

The second scenario for imposing measurement at a specific phase is to use a method of
state discrimination other than frequency measurement.  Selection rules can be used to
obtain a different response from the two atomic states, $|0\rangle$ and $|1\rangle$,
to the detection laser. In Fig.~\ref{f.atomlevels}(c) energy levels are illustrated
for a ground and excited state manifold that have the same angular momentum, $F'=F
(=1/2$ in the illustration). If circularly polarized light is used to drive a $\sigma
+$ detection transition, only the $|0\rangle$ to $|2\rangle$ transition can be
excited; $|1\rangle$ does not couple to any excited state for this polarization. This
method of state discrimination is not subject to the Fourier transform limit for
measuring time--the existence or absence of scattered photons is sufficient to
determine the atomic state, and therefore the measurement time can be much shorter
than $\tau_\phi$. The magnetic field required to provide the quantization axis
determines the frequency splitting of the two-state system and therefore $\tau_\phi$.
The size of the field is limited only by the necessity to prevail over any stray
fields that may be present and can be quite small, easily enabling $\tau_m \ll
\tau_\phi$. For example, for the $^{87}$Rb transition at 795~nm, $\tau_m \sim \tau_2
\sim 10~$ns, while $\tau_\phi$ is 100~$\mu$s for $B=10$~mG. The drawback of this
approach is that the detection transition is not closed, and after scattering a small
number of photons, the atom is optically pumped to $|1\rangle$, where it is ``dark''
to the detection drive. For equal Clebsch-Gordan coefficients for the transitions
involved, the number of photons scattered before the atom is pumped into the dark
state can be seen to be $\sum_{n=1}^{\infty}n2^{-n}=2$. This is in contrast to
thousands of photons that can easily be scattered on a closed transition without
repump light.

\section{Summary}

In conclusion, we have discussed the possibility that the relative phase in quantum
mechanics plays a role in measurement outcome.  The value of the relative phase at the
moment of measurement naturally averages out over an ensemble, and the difficulty in
imposing a measurement at a specific $\phi$ likely signifies that experiments have not
been sensitive to any possible effect.  We have presented two specific scenarios for
forcing a measurement to occur at a specific phase, each of which should be achievable
in current cold-atom systems.



\begin{thebibliography}{}
%
%

\bibitem{wheeler}
Wheeler,~J.~A., Zurek,~W.~H. (eds.): Quantum Theory and Measurement. Princeton
University Press, Princeton, NJ (1983)

\bibitem{adler}
Adler,~S.~L.: Probability in Orthodox Quantum Mechanics: Probability as a Postulate
Versus Probability as an Emergent Phenomenon, quant-ph/0004077 (2000)

\bibitem{smolin}
Smolin, L.: The Trouble with Physics, pp.~3--11. Houghton Mifflin Company, Boston
(2007)

\bibitem{MW}
Everett~III,~H.: ``Relative State'' Formulation of Quantum Mechanics. Rev. Mod. Phys.
29, 454--462 (1957)

\bibitem{histories}
Griffiths,~R.~B.: Consistent Quantum Theory. Cambridge University Press, Cambridge,
England (2002)

\bibitem{CSL1}
Ghirardi,~G.~C., Pearle,~P., Rimini,~A.: Markov processes in Hilbert space and
continous spontaneous localization of systems of identical particles. Phys. Rev. A 42,
78--89 (1990)

\bibitem{CSL2}
Adler,~S.~L., Bassi,~A.: Is Quantum Theory Exact? Science 325, 275--276 (2009)

\bibitem{dBB}
Bohm,~D.: A Suggested Interpretation of the Quantum Theory in Terms of ``Hidden''
Variables. Phys. Rev. 85, 166--179 (1952)

\bibitem{mod}
We implicitly mean a specific value of $\phi~{\rm mod}~2\pi$.

\bibitem{bell}
Bell,~J.~S.: On the Einstein-Podolsky-Rosen Paradox. Physics 1, 195--200 (1964).

\bibitem{interferometer}
Godun,~R.~M., D'Arcy,~M.~B., Summy,~G.~S., Burnett,~K.: Prospects for atom
interferometry. Contemporary Physics 42, 77--95 (2001)

\bibitem{herzberg}
See, for example, Herzberg,~G.: Atomic Spectra and Atomic Structure, pp.~82--119.
Dover, New York (1944)

\bibitem{cohen-tann}
Cohen-Tannoudji,~C., Dupont-Roc,~J., Grynberg,~G.: Atom-Photon Interactions,
pp.~259--266. John Wiley and Sons, Inc., New York (1992)

\bibitem{null}
See, for example, Porrati,~M., Putterman~S.: Wave-function collapse due to null
measurements: The origin of intermittent atomic fluorescence. Phys. Rev. A 36,
929--932 (1987), for a discussion of state reduction via null measurement.

\bibitem{meas_time}
More precisely, this could be called the minimum measurement time or the quantum
measurement time.  In the presence of technical noise, more averaging may be required
to determine the state of the system.

\bibitem{eom}
Mueller,~H., Chiow,~S., Herrmann,~S., Chu,~S.: Nanosecond electro-optical switching
with a repetition rate above 20~MHz. Rev. Sci. Inst. 78, 124702 (2007)

\end{thebibliography}


\end{document}